\def\ra{\rangle}

\def\be{\begin{equation}}
\def\ee{\end{equation}}
\def\ba{\begin{array}}
\def\ea{\end{array}}

\def\Cb{{\Bbb C}}

\documentclass[aps,pra,showpacs,showkeys,preprint]{revtex4}
\usepackage{amssymb}
\usepackage{amsfonts}
\usepackage{mathrsfs}
\usepackage{amssymb}
\usepackage{mathrsfs}
\usepackage{amsmath}
\usepackage{graphicx}
\begin{document}
\baselineskip=20pt
\title{Representation Class and Geometrical Invariants of Quantum States
under Local Unitary Transformations}

\author{Zu-Huan Yu$^{*}$}
\author{Xian-Qing Li-Jost$^{\dagger}$}
\author{Shao-Ming Fei$^{*,\ \dagger}$}

\affiliation{$^{*}$ School of Mathematical Sciences, Capital Normal
University, Beijing 100037, China\\
$^{\dagger}$ Max Planck Institute for Mathematics in the Sciences,
D-04103 Leipzig, Germany}

\begin{abstract}
We investigate the equivalence of bipartite quantum mixed states under local
unitary transformations by introducing representation classes from a geometrical approach.
It is shown that two bipartite mixed states are
equivalent under local unitary transformations if and only if they have the
same representation class. Detailed examples are given on calculating
representation classes.
\end{abstract}

\pacs{03.67.-a, 02.20.Hj, 03.65.-w}
\keywords{Schmidt decomposition, Bipartite mixed state, Local
invariant, Representation class }
\maketitle


\section{Introduction}

As a key physical resource in quantum information processing
such as quantum cryptography, quantum
teleportation and quantum computation \cite{nielsen},
quantum entanglement has been recently extensively investigated.
Due the fact that the properties of entanglement for multipartite quantum systems
remain invariant under local unitary transformations on the individual subsystems,
the entanglement can be characterized in principle by all the invariants
under local unitary transformations. For instance,
the trace norms of realigned or partial transposed density
matrices in entanglement measure \cite{prl2} and separability criteria \cite{chen}
are some of these invariants. Therefore a complete set of invariants
gives rise to the classification of the quantum states under local
unitary transformations. Two quantum states are locally
equivalent if and only if all these invariants have equal
values for these states.

There are many ways to construct such invariants
of local unitary transformations.
The method developed in \cite{Rains,Grassl}, in principle, allows
one to compute all the invariants of local unitary
transformations, though it is not generally operational.
There have been some results on calculating invariants
related to the equivalence of quantum states under local unitary transformations,
e.g. for general two-qubit systems \cite{Makhlin},
three-qubit states \cite{Linden99,sun3qubit},
some generic mixed states \cite{generic,goswami,sungeneric},
tripartite pure and mixed states \cite{wl}.
In particular, in terms of the Bloch representation of
density matrices for general two-qubit systems,
18 invariants have been presented in \cite{Makhlin}. It has been shown that these 18 invariants
are sufficient to guarantee that two two-qubit states are equivalent
under local unitary transformations, and lack of anyone of these
18 invariants would result in incompleteness of the set of invariants.

However generally we still have no operational criteria to judge the equivalence for two
general bipartite mixed states under local unitary transformations.
In this letter we investigate the equivalence of quantum states under
local unitary transformations according to the spectral decompositions
and the Schmidt expressions of the eigenvectors of bipartite density matrices.
We give a general theorem on the local equivalence relations.
From this theorem one can in principle construct the complete set of invariants
under local unitary transformations, according to detailed cases.
For comparison we calculate the invariants for two-qubit systems.
Marvelously in our scheme we only need at most 12 invariants to
characterize the local equivalence of two-qubit systems. As an example
we also study in detail the invariants for qubit-qutrit systems.

\section{Representation class and geometrical invariants}

Let $\mathcal {H}_{1}$ and $\mathcal {H}_{2}$ be complex Hilbert
spaces of dimension $m$ and $n$ respectively, $m\geq n\geq 2$.
The tensor space $\mathcal {H}=\mathcal {H}_{1}\otimes \mathcal {H}_{2}$ is a
complex Hilbert space of dimension $mn$.
Let $\rho$ and $\tilde{\rho}$ be two bipartite density
matrices defined on $\mathcal {H}=\mathcal {H}_{1}\otimes \mathcal {H}_{2}$.
$\rho$ and $\tilde{\rho}$ are said to be equivalent
under local unitary transformations if there exist unitary operators
$U$ on $\mathcal {H}_{1}$ and $V$ on $\mathcal {H}_{2}$ such that
\be\label{eq}
\tilde{\rho}=(U\otimes V)\rho(U\otimes V)^\dag,
\ee
where $\dag$ stands for transpose and conjugation.

As a hermitian operator, a mixed state $\rho$ with rank $l$ has the spectral decomposition
\be\label{rho}
\rho=\lambda_{1}|e_{1}\rangle \langle e_{1}|+\cdots +\lambda_{l}
|e_{l}\rangle \langle e_{l}|,
\ee
where $\lambda_{i}$, $i=1,\cdots,l$, are the nonzero eigenvalues of $\rho$,
$|e_{i}\rangle$ are the corresponding eigenvectors associated with $\lambda_{i}$, which can be chosen
as orthonormal vectors. For convenience we set
$\lambda_{1} \geq \lambda_{2} \geq \cdots \geq \lambda_{l}>0$.

Every eigenvector $|e_{i}\rangle$ with Schmidt rank $k_i$ has Schmidt decomposition, namely
there exist orthonormal vectors $a_{j}^{i}$ and $b_{j}^{i}$
of $\mathcal{H}_{1}$ and $\mathcal {H}_{2}$ respectively, $j=1,\cdots,
k_{i}$, such that
$$|e_{i}\rangle=\mu_{i}^{1}a_{1}^{i}\otimes b_{1}^{i}+\cdots
+\mu_{i}^{k_{i}}a_{k_{i}}^{i}\otimes b_{k_{i}}^{i},\ \ k_{i} \leq n,
\ \ i=1,\cdots,l,$$
where $\mu_{i}^{j}$, $j=1,\cdots, k_{i}$, are
so called Schmidt coefficients satisfying
$(\mu_{i}^{1})^{2}+ \cdots +(\mu_{i}^{k_{i}})^{2}=1$. Without loss of
generality we assume $\mu_{i}^{1}\geq \mu_{i}^{2}\geq \cdots \geq
\mu_{i}^{k_{i}}>0.$

We extend the set of $k_1$ orthonormal vectors $a_{1}^{1}, a_{2}^{1}, \cdots, a_{k_{1}}^{1}$ to be an
orthonormal basis of $\mathcal {H}_{1}$, $\{a_{1}, a_{2}, \cdots , a_{k_{1}}, \cdots , a_{m}\}$,
and $b_{1}^{1}, b_{2}^{1},\cdots, b_{k_{1}}^{1}$ to an orthonormal basis of $\mathcal {H}_{2}$,
$\{b_{1}, b_{2}, \cdots , b_{k_{1}}, \cdots , b_{n}\}$.
Therefore the vectors $a_{j}^{i}$ and $b_{j}^{i}$, $j=1,\cdots, k_{i}$ can be
represented according to the two bases respectively,
\be\label{XY}
(a_{1}^{i}, a_{2}^{i}, \cdots, a_{k_{i}}^{i})=(a_{1}, a_{2}, \cdots ,
a_{m})X_{i},\ \ \ (b_{1}^{i}, b_{2}^{i}, \cdots,
b_{k_{i}}^{i})=(b_{1}, b_{2}, \cdots , b_{n})Y_{i},
\ee
for some $m\times k_{i}$ matrix $X_{i}$ and $n\times k_{i}$ matrix $Y_{i}$. Denote
$r(\rho)_{i}=(\lambda_{i}, \mu_{i}^{1},\cdots,
\mu_{i}^{k_{i}},X_{i},Y_{i})$, $i=1,...,l$.
We say that
\be\label{rc}
r(\rho)=(r(\rho)_{1}, \cdots,r(\rho)_{l})
\ee
is  a representation of the
mixed state $\rho$. We call the set of all the representations of $\rho$
the representation class of $\rho$, denoted by $\mathscr{R}(\rho)$.

{\bf [Theorem]} Two mixed states $\rho$ and $\tilde{\rho}$ of
bipartite quantum systems are equivalent under local unitary transformations
if and only if they have the same
representation class, i.e. $\mathscr{R}(\rho)=\mathscr{R}(\tilde{\rho}).$

{\bf [Proof]} We first prove the sufficient part of the condition.
Assume that the mixed states $\rho$ and $\tilde{\rho}$ have the same
representation class, $\mathscr{R}(\rho)=\mathscr{R}(\tilde{\rho}).$
Hence there exists a representation $r(\rho)\in \mathscr{R}(\rho)$,
$r(\tilde{\rho})\in \mathscr{R}(\tilde{\rho})$ such that
$r(\rho)=r(\tilde{\rho})$. Let us assume that
$a_{1},\cdots, a_{m}$ and $b_{1},\cdots, b_{n}$ be the orthonormal basis of
$\mathcal {H}_{1}$ and $\mathcal {H}_{2}$ with respect to the
representation $r(\rho)$, ${\tilde {a}}_{1},\cdots, {\tilde {a}}_{m}$ and
${\tilde{b}}_{1},\cdots, {\tilde {b}}_{n}$ the orthonormal basis of $\mathcal {H}_{1}$ and
$\mathcal {H}_{2}$ with respect to the representation $r(\tilde{\rho})$.

Then there exist unitary transformations $U$ on $\mathcal {H}_{1}$ and $V$
on $\mathcal {H}_{2}$ such that
$$
({\tilde {a}}_{1},\cdots,{\tilde{a}}_{m})=U(a_{1},\cdots,a_{m}),\
\ ({\tilde {b}}_{1},\cdots,{\tilde{b}}_{n})=V(b_{1},\cdots,b_{n}).
$$
From $r(\rho)=r(\tilde{\rho})$ we have
$\lambda_{i}={\tilde {\lambda}}_{i}$, $\mu_{i}^{j}={\tilde
{\mu}}_{i}^{j}$, $X_{i}={\tilde {X}}_{i}$, $Y_{i}={\tilde
{Y}}_{i}$ for $i=1,\cdots, l$, $j=1,\cdots, k_{i}$. Therefore
$\tilde{\rho}=(U\otimes V)\, \rho\, (U^{\dagger}\otimes
V^{\dagger})$.

For the necessary pat of the condition, we assume
$\tilde {\rho}=(U\otimes V) \, \rho \, (U\otimes
V)^{\dagger}$. If $r(\rho)$ is a representation of $\rho$, from the
spectral decomposition of mixed states, and the properties of the
unitary transformations, we have that $r(\rho)$ is also a
representation of $\tilde {\rho}$, hence $\mathscr{R}(\rho)\subset
\mathscr{R}(\tilde {\rho})$. Similarly, we have $\mathscr{R}(\tilde
{\rho})\subset \mathscr{R}(\rho)$. Therefore
$\mathscr{R}(\rho)=\mathscr{R}(\tilde {\rho})$.
\hfill $\Box$

{\bf {Remark}}\quad The representations in the representation class of a mixed state
are not independent. Actually if one representation is given, the others are also known.
And the equivalence of two quantum states can be studied by calculating their
representation classes.

We consider as an example the two-qubit systems, $m=n=2$.
Generally, a mixed state $\rho$ has four different
eigenvalues $\lambda_{i}, (i=1,\cdots,4)$. Here as the trace of $\rho$ is
one, only three eigenvalues are independent. Let
$|e_{i}\rangle$, $i=1,\cdots,4$, be the corresponding orthonormal
eigenvectors. Since $|e_{4}\rangle$ is determined by other three
eigenvectors up to a scale $e^{i\theta}$, we only need to take into account
three eigenvectors. Every eigenvector of these three
$|e_{i}\rangle$, $i=1,2,3$, can have at most Schmidt rank two.
But only one of the Schmidt coefficients $\mu_{i}^{1}$, $\mu_{i}^{2}$
of $|e_{i}\rangle$, $i=1,2,3$, is independent. Therefore only three
eigenvalues and three Schmidt coefficients (all together 6 quantities) are
free. The matrices $X_{1}$ and $Y_{1}$ are unit matrices of
order $2$. While $X_{2},Y_{2}$, $X_{3},Y_{3}$ are unitary matrices of
order $2$, taking the following form
$$
\left(\begin{array}{cccc} re^{i
\alpha_{1}} &
-\sqrt{1-r^{2}}e^{-i \alpha_{2}}e^{i\alpha_{3}}\\
\sqrt{1-r^{2}}e^{i\alpha_{2}} & re^{-i \alpha_{1}}e^{i\alpha_{3}}
\end{array}\right),
$$
where $r>0$, $\alpha_1,\alpha_3,\alpha_3\in \mathbb{R}$. That is,
every matrix has four free quantities. Since $|e_{i}\rangle$,
$i=1,2,3$, are perpendicular to each other, and $|e_{2}\rangle$ and
$|e_{3}\rangle$ are determined up to a phase factor $e^{i\theta}$, there are only
$6$ free parameters left. Therefore we only need at most 12
invariants to check the local equivalence for two-qubit bipartite
quantum systems, which is different from \cite{Makhlin} where 18
invariants are needed.

As an example we consider a two-qubit modified Werner state
\be\label{werner}
\rho=\left(
\ba{cccc}
(1-e-f)/3&0&0&0\\
0&(1+2f)/6&(1-4f)/6&0\\
0&(1-4f)/6&(1+2f)/6&0\\
0&0&0&(1+e-f)/3
\ea
\right),
\ee
where $0\leq f\leq 1-e$, $e\geq 0$.
When $e=0$, $\rho$ is just the usual two-qubit Werner state \cite{werner}, which is separable
for $f\leq 1/2$.

$\rho$ has eigenvalues $\lambda_1=(1-f+e)/3$,
$\lambda_2=(1-f)/3$, $\lambda_3=(1-f-e)/3$,
$\lambda_4=f$, with the corresponding eigenvectors
$$
|\nu_1>=\left(\ba{c} 0\\0\\0\\1\ea\right),~~
|\nu_2>=\frac{1}{\sqrt{2}}\left(\ba{c} 0\\1\\1\\0\ea\right),~~
|\nu_3>=\left(\ba{c} 1\\0\\0\\0\ea\right),
~~|\nu_4>=\frac{1}{\sqrt{2}}\left(\ba{c} 0\\-1\\1\\0\ea\right).
$$

Set $a_1=b_1=(0,1)^t$, $a_2=b_2=(1,0)^t$, we have $|\nu_1>=a_1\otimes b_1$.
From (\ref{XY}), we have $X_1=Y_1=(1,0)^t$.
Up to a global phase factor, the eigenvector $|\nu_2>$ can be expressed as
$$
|\nu_2>=\frac{1}{\sqrt{2}}(a_1\otimes b_2+a_2\otimes b_1)
\sim \frac{1}{\sqrt{2}}e^{i\theta}(e^{i\theta_2}a_1\otimes e^{-i\theta_2}b_2+e^{i\theta_1}a_2\otimes e^{-i\theta_1}b_1)
=\frac{1}{\sqrt{2}}(a_1^2\otimes b_1^2+a_2^2\otimes b_2^2),
$$
where $a_1^2=e^{i\theta}e^{i\theta_2}a_1$, $a_2^2=e^{i\theta}e^{i\theta_1}a_2$,
$b_1^2=e^{-i\theta_2}b_2$, $b_2^2=e^{-i\theta_1}b_1$.
From (\ref{XY}), we have
$$
X_2=e^{i\theta}\left( \ba{cc}e^{i\theta_2}&0\\0&e^{i\theta_1}\ea \right),~~~
Y_2=\left( \ba{cc}0&e^{-i\theta_1}\\e^{-i\theta_2}&0\ea \right).
$$
Similarly one can obtain
$$
X_3=\left( \ba{c}0\\e^{i(\beta+\beta_1)}\ea \right),~~
Y_3=\left( \ba{c}0\\e^{-i\beta_1}\ea \right),~~
X_4=e^{i\gamma}\left( \ba{cc}e^{i\gamma_2}&0\\0&-e^{i\gamma_1}\ea \right),~~
Y_4=\left( \ba{cc}0&e^{-i\gamma_1}\\e^{-i\gamma_2}&0\ea \right).
$$

The representation class $\mathscr{R}(\rho)$
is given by $\mathscr{R}(\rho)=(r(\rho)_1,r(\rho)_2,r(\rho)_3,r(\rho)_4)$, where
\be\label{rpwerner}
\ba{l}
\displaystyle r(\rho)_1=(\frac{1-e-f}{3},1,X_1,Y_1),~~~
r(\rho)_2=(\frac{1-f}{3},\frac{1}{\sqrt{2}},\frac{1}{\sqrt{2}},X_2,Y_2),\\[4mm]
\displaystyle r(\rho)_3=(\frac{1+e-f}{3},1,X_3,Y_3),~~~
r(\rho)_4=(f,\frac{1}{\sqrt{2}},\frac{1}{\sqrt{2}},X_4,Y_4).
\ea
\ee
This representation class is parameterized by 8 free parameters.
Any states with representations of the form (\ref{rpwerner}),
for some given values $\theta,\theta_1,\theta_2,\gamma,\gamma_1,\gamma_2,\beta,\beta_1$,
are equivalent to the state (\ref{werner}) under local unitary transformations.

\section{Representation class for qubit-qutrit systems}

The representation class can be analytically calculated in principle according to
detailed situations: the rank of the density matrix, the property of the
eigenvalues and the Schmidt ranks of the eigenvectors.
In the following we investigate as another example in detail
the local equivalence of qubit-qutrit systems, $m=2$, $n=3$.
We compute the representations for the cases that the mixed states have two
different nonzero eigenvalues.

Let $\lambda_{1}$ and $\lambda_{2}$ be the two nonzero eigenvalues
of $\rho$, $\lambda_{1}>\lambda_{2}$, $\lambda_{1}+\lambda_{2}=1$, with
$|e_{1}\rangle$ and $|e_{2}\rangle$ the corresponding
eigenvectors. Then mixed state $\rho$ has the following spectral
decomposition
$$
\rho=\lambda_{1}|e_{1}\rangle \langle e_{1}|+\lambda_{2}
|e_{2}\rangle \langle e_{2}|.
$$
As the eigenvalues are
different, for given $\rho$, $|e_{j}\ra$ is determined up to a
phase factor $e^{i\theta_{j}}$. We calculate the representation classes according to
various detailed cases.

{\bf Case 1} \quad The eigenvectors $|e_{1}\rangle$ and $|e_{2}\rangle$ are all separable,
i.e. the Schmidt rank $k_{1}=k_{2}=1$. Their Schmidt decompositions are of the forms
$$
|e_{1}\ra=a^{1}_{1}\otimes b^{1}_{1},  \quad \quad |e_{2}\ra=a^{2}_{1}\otimes
b^{2}_{1},
$$
where for fixed $\rho$, $a_{i}$ and $b_{i}$ are determined up to a phase factor $e^{i\theta}$.
To calculate the matrices $X_i$, $Y_i$ in (\ref{XY}) we choose the orthonormal basis of $\mathcal {H}_{1}$
to be $\{a_{1}, a _{2}\}$ with $a_{1}=a^{1}_{1}$, which is determined up to a rotation
$$
\left(\begin{array}{cccc}e^{i \alpha_{1}} & 0\\
0 & e^{i \alpha_{2}}
\end{array}\right),~~~~\alpha_{1},~\alpha_{2}\in\mathbb{R}
$$
and the orthonormal basis of $\mathcal {H}_{2}$ to be $\{b_{1}, b _{2}, b_{3}\}$ with
$b_{1}=b^{1}_{1}$, up to a rotation
$$
\left(\begin{array}{cccc}e^{i
\alpha} & 0\\
0 & u(2)
\end{array}\right),
$$
where $\alpha\in\mathbb{R}$, $u(2)\in U(2)$ is a $2\times 2$ unitary
matrix.

Therefore according to (\ref{XY}) if we set $a^{1}_{1}=(a_{1},
a_{2})X_{1}$, $b^{1}_{1}=(b_{1}, b_{2}, b_{3})Y_{1}$, and
$a^{2}_{1}=(a_{1}, a_{2})X_{2}$, $b^{2}_{1}=(b_{1}, b_{2},
b_{3})Y_{2}$, we have
\be\label{x1y1}
X_{1}=\left(
          \begin{array}{c}
            1 \\
            0 \\
          \end{array}
        \right),\quad
Y_{1}=\left(
                         \begin{array}{c}
                           1 \\
                           0 \\
                           0 \\
                         \end{array}
                       \right);\quad \quad
X_{2}=\left(
          \begin{array}{cc}
            e^{i\theta_{1}} & 0 \\
            0 & e^{i\theta_{2}} \\
          \end{array}
        \right)X^{0}_{2},\quad
        Y_{2}=\left(
          \begin{array}{cc}
            e^{i\theta} & 0 \\
            0 & u(2) \\
          \end{array}
        \right)Y^{0}_{2},
\ee
where
\be\label{pp}
                       X^{0}_{2}=\left(
          \begin{array}{c}
            x_{1} \\
            x_{2} \\
          \end{array}
        \right),\quad
        Y^{0}_{2}=\left(
                         \begin{array}{c}
                           y_{1} \\
                           y_{2} \\
                           y_{3} \\
                         \end{array}
                       \right)
\ee
for some $x_1,x_2$ and $y_1,y_2,y_3\in\Cb$.
The representation of $\rho$ is given by
$r(\rho)=(r(\rho)_{1}, r(\rho)_{2})$, where
$r(\rho)_{1}=(\lambda_{1}, 1, X_{1}, Y_{1}),$ $r(\rho)_{2}=(\lambda_{2}, 1, X_{2}, Y_{2})$.
The representation class is
 $$\mathscr{R}( \rho)=\{(r(\rho)_{1}, r(\rho)_{2}) \ | \ \theta_{i}\in
 \mathbb{R},\ u(2)\in U(2)\}.$$

{\bf Case 2}  \quad The eigenvector $|e_{1}\rangle$ is separable, while
$|e_{2}\rangle$ is entangled, i.e. $k_{1}=1, \ k_{2}=2$. The Schmidt
decompositions are of the forms
$$
|e_{1}\rangle=a^{1}_{1}\otimes b^{1}_{1}, \ \
|e_{2}\rangle=\mu^{1}_{2}\,a^{2}_{1}\otimes
b^{2}_{1}+\mu^{2}_{2}\,a^{2}_{2}\otimes b^{2}_{2}.
$$
We choose the orthonormal basis of $\mathcal {H}_{1}$ (resp. $\mathcal {H}_{2}$) to be
$\{a^{1}_{1}, a _{2}\}$ (resp. $\{b^{1}_{1}, b _{2}, b_{3}\}$) as defined in the case 1.
Then we have the same $X_1$ and $Y_1$ as in (\ref{x1y1}). Set
\be\label{x02y02}
X^{0}_{2}=\left(
          \begin{array}{ccc}
            x_{1} & x_{2} \\
            x_{3} & x_{4} \\
          \end{array}
        \right),\quad  Y^{0}_{2}=\left(
                         \begin{array}{cccc}
                           y_{1} & y_{2}  \\
                           y_{3} & y_{4} \\
                           y_{5} & y_{6}  \\
                         \end{array}
                       \right).
\ee
If $\mu^{1}_{2}\neq \mu^{2}_{2}$, then $a^{2}_{1}, \  b^{2}_{1};\
a^{2}_{2},\ b^{2}_{2}$ are determined up to a factor $e^{i\theta}$.
We have
$$X_{2}=\left(
          \begin{array}{cc}
            e^{i\theta_{1}} & 0 \\
            0 & e^{i\theta_{2}} \\
          \end{array}
        \right)X^{0}_{2}\left(
          \begin{array}{cc}
            e^{i\beta_{1}} & 0 \\
            0 & e^{i\beta_{2}} \\
          \end{array}
        \right),\quad  Y_{2}=\left(
          \begin{array}{cc}
            e^{i\theta} & 0 \\
            0 & u(2) \\
          \end{array}
        \right)Y^{0}_{2}\left(
          \begin{array}{cc}
            e^{i\alpha_{1}} & 0 \\
            0 & e^{i\alpha_{2}} \\
          \end{array}
        \right),
        $$
where $\beta_{1}+\alpha_{1}=\beta_{2}+\alpha_{2}$.
The representations of $\rho$ are $r(\rho)=(r(\rho)_{1},
r(\rho)_{2})$ with
 $r(\rho)_{1}=(\lambda_{1}, 1, X_{1}, Y_{1}),$ $r(\rho)_{2}=(\lambda_{2}, \mu^{1}_{2}, \mu^{2}_{2}, X_{2}, Y_{2})$.
The representation class is given by
 $$\mathscr{R}( \rho)=\{(r(\rho)_{1}, r(\rho)_{2}) \ | \ \theta, \beta _{i}, \theta_{i}\in
 \mathbb{R},\beta_{1}+\alpha_{1}=\beta_{2}+\alpha_{2},\ u(2)\in U(2)\}.$$

If $\mu^{1}_{2}=\mu^{2}_{2}=\frac{\sqrt{2}}{2}$, then $a^{2}_{1}, \
b^{2}_{1}; \ a^{2}_{2},\ b^{2}_{2}$ are determined up to a rotation
$u(2)$. The representations of $\rho$ are $r(\rho)=(r(\rho)_{1},
r(\rho)_{2})$ with
 $r(\rho)_{1}=(\lambda_{1}, 1, X_{1}, Y_{1}),$ $r(\rho)_{2}=(\lambda_{2}, \frac{\sqrt{2}}{2}, \frac{\sqrt{2}}{2}, X_{2}, Y_{2}),$
where
$$
X_{2}=\left(
          \begin{array}{cc}
            e^{i\theta_{1}} & 0 \\
            0 & e^{i\theta_{2}} \\
          \end{array}
        \right)X^{0}_{2}\ u_{1}(2),\quad  Y_{2}=\left(
          \begin{array}{cc}
            e^{i\theta} & 0 \\
            0 & u_{2}(2) \\
          \end{array}
        \right)Y^{0}_{2}\ {u}_{1}(2)^\dag,~~~u_{1}(2),~ u_{2}(2)\in U(2).
$$
The representation class is given by
 $$\mathscr{R}( \rho)=\{(r(\rho)_{1}, r(\rho)_{2}) \ | \ \theta, \theta_{i}\in
 \mathbb{R};\  u_{1}(2), u_{2}(2)\in U(2)\}.$$

{\bf Case 3} \quad The eigenvector $|e_{2}\rangle$ is separable, while $|e_{1}\rangle$
is entangled, i.e. $k_{1}=2, \ k_{2}=1$. The Schmidt decompositions are of the forms
$$
|e_{1}\rangle=\mu^{1}_{1}\,a^{1}_{1}\otimes
b^{1}_{1}+\mu^{2}_{1}\,a^{1}_{2}\otimes b^{1}_{2}, \ \
|e_{2}\rangle=a^{2}_{1}\otimes b^{2}_{1}.
$$
In the bases $\{a^{1}_{1}, a^{1}_{2}\}$ (resp. $\{b^{1}_{1}, b^{1}_{2}, b_{3}\}$)
of $\mathcal {H}_{1}$ (resp. $\mathcal {H}_{2}$), at first we have
\be\label{x2y2}
X_{1}=\left(
           \begin{array}{cc}
             1 & 0 \\
             0 & 1 \\
           \end{array}
         \right), \ Y_{1}=\left(
                            \begin{array}{cc}
                              1 & 0 \\
                              0 & 1 \\
                              0 & 0 \\
                            \end{array}
                          \right).
\ee
If $\mu^{1}_{1}\neq \mu_{1}^{2}$, the representations of $\rho$ are
$r(\rho)=(r(\rho)_{1}, r(\rho)_{2})$ with
 $r(\rho)_{1}=(\lambda_{1}, \mu^{1}_{1}, \mu^{2}_{1}, X_{1}, Y_{1}),$
$r(\rho)_{2}=(\lambda_{2}, 1, X_{2}, Y_{2}),$ where
$$X_{2}=\left(
          \begin{array}{cc}
            e^{i\theta_{1}} & 0 \\
            0 & e^{i\theta_{2}} \\
          \end{array}
        \right)X^{0}_{2},\quad  Y_{2}=\left(
          \begin{array}{ccc}
            e^{i\beta_{1}} & 0 &0\\
            0 & e^{i\beta_{2}}&0 \\
0 & 0& e^{i\beta_{3}}\\
          \end{array}
        \right)Y^{0}_{2},$$
with $\theta_{1}+\beta_{1}=\theta_{2}+\beta_{2}$ and $X^{0}_{2}$ and $Y^{0}_{2}$ given
in (\ref{pp}). The representation class is
$$
\mathscr{R}( \rho)=\{(r(\rho)_{1}, r(\rho)_{2}) \ | \ \beta_{i}, \theta_{i}\in
 \mathbb{R};\ \theta_{1}+\beta_{1}=\theta_{2}+\beta_{2}  \}.
$$

If $\mu^{1}_{1}=\mu_{1}^{2}=\frac{\sqrt{2}}{2}$, the representations
of $\rho$ are given by $r(\rho)=(r(\rho)_{1}, r(\rho)_{2})$ with
 $r(\rho)_{1}=(\lambda_{1}, \frac{\sqrt{2}}{2},$ $ \frac{\sqrt{2}}{2}, X_{1}, Y_{1}),$
$r(\rho)_{2}=(\lambda_{2}, 1, X_{2}, Y_{2}),$ where
$$X_{2}=u(2)X^{0}_{2},\quad  Y_{2}=\left(
          \begin{array}{ccc}
            u(2)^\dag & 0 \\
            0 & e^{i\beta} \\
          \end{array}
        \right) Y^{0}_{2}\, e^{i\theta}.$$
The representation class is
 $$\mathscr{R}( \rho)=\{(r(\rho)_{1}, r(\rho)_{2}) \ | \ \beta, \theta\in
 \mathbb{R};\ u(2)\in U(2)\}.$$

{\bf Case 4}\quad Both eigenvectors $|e_{1}\rangle$ and $|e_{2}\rangle$ are entangled,
i.e. $k_{1}=k_{2}=2$. Their Schmidt decompositions are given by
$$|e_{1}\rangle=\mu_{1}^{1}\,a^{1}_{1}\otimes b^{1}_{1}+\mu_{1}^{2}\,a^{1}_{2}\otimes b^{1}_{2}, \ \
|e_{2}\rangle=\mu^{1}_{2}\,a^{2}_{1}\otimes
b^{2}_{1}+\mu^{2}_{2}\,a^{2}_{2}\otimes b^{2}_{2}.$$

In the basis $\{a^{1}_{1}, a^{1}_{2}\}$ (resp. $\{b^{1}_{1}, b^{1}_{2}, b_{3}\}$)
of $\mathcal {H}_{1}$ (resp. $\mathcal {H}_{2}$), we have $X_1$ and $Y_1$ as given in
(\ref{x2y2}).
If $\mu^{1}_{1}>\mu^{2}_{1},\ \mu^{1}_{2}>\mu^{2}_{2}$, we have
$$X_{2}=\left(
          \begin{array}{cc}
            e^{i\theta_{1}} & 0 \\
            0 & e^{i\theta_{2}} \\
          \end{array}
        \right)X^{0}_{2}\left(
          \begin{array}{cc}
            e^{i\gamma_{1}} & 0 \\
            0 & e^{i\gamma_{2}} \\
          \end{array}
        \right),\quad  Y_{2}=\left(
          \begin{array}{ccc}
            e^{i\beta_{1}} & 0 &0\\
            0 & e^{i\beta_{2}}&0 \\
0 & 0& e^{i\beta_{3}}\\
          \end{array}
        \right)Y^{0}_{2}\left(
          \begin{array}{cc}
            e^{i\alpha_{1}} & 0 \\
            0 & e^{i\alpha_{2}} \\
          \end{array}
        \right),$$
with $\theta_{1}+\beta_{1}=\theta_{2}+\beta_{2}$,
$\gamma_{1}+\alpha_{1}=\gamma_{2}+\alpha_{2}$, where
$X^{0}_{2}$ and $Y^{0}_{2}$ are given in (\ref{x02y02}).
Hence $r(\rho)_{1}=(\lambda_{1}, \mu^{1}_{1}, \mu^{2}_{1},
X_{1}, Y_{1})$ and $r(\rho)_{2}=(\lambda_{2}, \mu^{1}_{2}, \mu^{2}_{2},
X_{2}, Y_{2})$. And the representation class is
$$
\mathscr{R}( \rho)=\{(r(\rho)_{1}, r(\rho)_{2}) \ | \ \beta_{i}, \theta_{i}, \gamma_{i}, \alpha_{i} \in
 \mathbb{R};\ \theta_{1}+\beta_{1}=\theta_{2}+\beta_{2},\
        \gamma_{1}+\alpha_{1}=\gamma_{2}+\alpha_{2}\}.
$$

If $\mu^{1}_{1}>\mu^{2}_{1},\
\mu^{1}_{2}=\mu^{2}_{2}=\frac{\sqrt{2}}{2}$, we have
$$X_{2}=\left(\begin{array}{cc}
            e^{i\theta_{1}} & 0 \\
            0 & e^{i\theta_{2}} \\
          \end{array}
        \right)X^{0}_{2}\,u(2),\quad  Y_{2}=\left(
          \begin{array}{ccc}
            e^{i\beta_{1}} & 0 &0\\
            0 & e^{i\beta_{2}}&0 \\
0 & 0& e^{i\beta_{3}}\\
          \end{array}
        \right)Y^{0}_{2}\,u(2)^\dag \,e^{i\theta},
$$
with $\theta_{1}+\beta_{1}=\theta_{2}+\beta_{2}$.
We have $r(\rho)_{1}=(\lambda_{1},
\mu^{1}_{1}, \mu^{2}_{1}, X_{1}, Y_{1})$ and $r(\rho)_{2}=(\lambda_{2},
\frac{\sqrt{2}}{2}, \frac{\sqrt{2}}{2}, X_{2}, Y_{2})$. The
representation class is
$$\mathscr{R}( \rho)=\{(r(\rho)_{1}, r(\rho)_{2}) \ | \ \theta, \theta_{i}, \beta_{i} \in
 \mathbb{R};\ \theta_{1}+\beta_{1}=\theta_{2}+\beta_{2},\ u(2)\in
 U(2)\}.$$

If $\mu^{1}_{1}=\mu^{2}_{1}=\frac{\sqrt{2}}{2},\
\mu^{1}_{2}>\mu^{2}_{2}$, we have
$$X_{2}=u(2)X^{0}_{2}\left(
          \begin{array}{cc}
            e^{i\theta_{1}} & 0 \\
            0 & e^{i\theta_{2}} \\
          \end{array}
        \right),\quad  Y_{2}=\left(
          \begin{array}{cc}
            u(2)^\dag & 0 \\
            0 & e^{i\theta} \\
          \end{array}
        \right)Y^{0}_{2}\left(
          \begin{array}{ccc}
            e^{i\beta_{1}} & 0\\
            0 & e^{i\beta_{2}} \\
          \end{array}
        \right),$$
with $\theta_{1}+\beta_{1}=\theta_{2}+\beta_{2}$.
In this case we have $r(\rho)_{1}=(\lambda_{1}, \frac{\sqrt{2}}{2}, \frac{\sqrt{2}}{2}, X_{1}, Y_{1})$
and $r(\rho)_{2}=(\lambda_{2}, \mu^{1}_{2}, \mu^{2}_{2}, X_{2}, Y_{2})$.
And the representation class is given by
$$\mathscr{R}( \rho)=\{(r(\rho)_{1}, r(\rho)_{2}) \ | \ \theta, \theta_{i}, \beta_{i} \in
 \mathbb{R};\ \theta_{1}+\beta_{1}=\theta_{2}+\beta_{2},\ u(2)\in
 U(2)\}.$$

If $\mu^{1}_{1}=\mu^{2}_{1}=\frac{\sqrt{2}}{2},\
\mu^{1}_{2}=\mu^{2}_{2}=\frac{\sqrt{2}}{2}$, we have
$$X_{2}=u_{1}(2)\,X^{0}_{2}\,u_{2}(2),\quad  Y_{2}=\left(
          \begin{array}{cc}
            {u}_{1}(2)^\dag & 0 \\
            0 & e^{i\theta} \\
          \end{array}
        \right)Y^{0}_{2}\,u_{2}(2)^\dag\, e^{i\gamma}
        $$
        and $r(\rho)_{1}=(\lambda_{1}, \frac{\sqrt{2}}{2}, \frac{\sqrt{2}}{2}, X_{1}, Y_{1}),$
$r(\rho)_{2}=(\lambda_{2}, \frac{\sqrt{2}}{2}, \frac{\sqrt{2}}{2},
X_{2}, Y_{2})$,
$$\mathscr{R}( \rho)=\{(r(\rho)_{1}, r(\rho)_{2}) \ | \ \theta, \gamma \in
 \mathbb{R};\ u_{1}(2), u_{2}(2) \in U(2)\}.$$

\section{Conclusion}

In stead of usual algebraic construction of invariants under local unitary transformations,
we have presented a geometrical approach to the classification of quantum bipartite
mixed states under local unitary transformations, which works for arbitrary $m\times n$ dimensional
quantum systems. It has been shown that two bipartite mixed states are
equivalent under local unitary transformations if and only if they have the
same representation class. As shown in the examples, these representation classes can be calculated in detail
according to the eigenvalues and the Schmidt decompositions of the eigenvectors
of density matrices. Although general analysis could be rather complicated as one has to take into
account many different cases, for a given density matrix the calculation would be quite direct.

{\bf Acknowledgements}  \quad The first author gratefully
acknowledge the support provided by China Scholarship Council and
Fund of Beijing Youxiurencai.
The work is partly supported by NSFC project
10675086, KM200510028022 and NKBRPC(2004CB318000).


\begin{thebibliography}{99}
\bibitem{nielsen} M.A. Nielsen and I.L. Chuang, Quantum Computation and
Quantum Information, Cambridge University Press, Cambridge, 2000.

\bibitem{prl2} K. Chen, S. Albeverio and S.M. Fei, Phys. Rev. Lett. \textbf{95}, 040504 (2005);\\
K. Chen, S. Albeverio and S.M. Fei, Phys. Rev. Lett. \textbf{95}, 210501 (2005).

\bibitem{chen}
K. Chen and L.A. Wu, Quant. Inf. Comput. \textbf{3}, 193 (2003).\\
K. Chen and L.A. Wu, Phys. Lett. A \textbf{306}, 14 (2002).\\
O. Rudolph, Physical Review A \textbf{67}, 032312 (2003).\\
K. Chen, S. Albeverio and S.M. Fei, Phys. Rev. A \textbf{68}, 062313 (2003).

\bibitem{Rains}
E.M.~Rains, {\em IEEE Transactions on Information Theory} \textbf{46} 54-59(2000).

\bibitem{Grassl}
M.~Grassl, M.~R\"otteler and T.~Beth, Phys. Rev. A \textbf{58}, 1833
(1998).

\bibitem{Makhlin} Y. Makhlin, Quan. Inf. Pro. \textbf{1}, 243 (2002).

\bibitem{Linden99} N. Linden, S. Popescu, and A. Sudbery, Phy. Rev. Lett.
\textbf{83}, 243 (1999).

\bibitem{sun3qubit}B.Z. Sun and S.M. Fei, Commun. Theor. Phys. {\bf 45}, 1007-1010 (2006).

\bibitem{generic} S. Albeverio, S.M. Fei, P. Parashar, W.L. Yang,
Phys. Rev. A \textbf{68}, 010303 (2003).

\bibitem{goswami}
S. Albeverio, S.M. Fei and D. Goswami,
Phys. Lett. A {\textbf 340} 37-42(2005).

\bibitem{sungeneric}
B.Z. Sun, S.M. Fei, X.Q. Li-Jost and Z.X. Wang,
J. Phys. A: Math. Gen. \textbf{39} L43-L47(2006).

\bibitem{wl}
S. Albeverio, L. Cattaneo, S.M. Fei and X.H. Wang,
Int. J. Quant. Inform. {\bf 3} 603-609 (2005);
Rep. Math. Phys. {\bf 56} 341-350 (2005).

\bibitem{werner} R.F. Werner, Phys. Rev. A \textbf{40}, 4277 (1989).
\end{thebibliography}
\end{document}